\documentclass[twocolumn,aps, prl, groupedaddress, eps]{revtex4}

\usepackage{graphicx}

\begin{document}

\title{Electric Field Modulation of Galvanomagnetic Properties of Mesoscopic Graphite}

\author{Yuanbo Zhang}
\author{Joshua P. Small}
\author{Michael E. S. Amori}
\author{Philip Kim}

\affiliation{Department of Physics and the Columbia Nanoscale Science
and Engineering Center, Columbia University, New York, New York 10027}

\begin{abstract}
Electric field effect devices based on mesoscopic graphite are
fabricated for galvanomagnetic measurements. Strong modulation of
magneto-resistance and Hall resistance as a function of gate
voltage is observed as sample thickness approaches
the screening length. Electric field dependent Landau level
formation is detected from Shubnikov de Haas oscillations in
magneto-resistance. The effective mass of electron and hole
carriers has been measured from the temperature dependant
behavior of these oscillations.

\end{abstract}

\maketitle


Graphite is a semimetal with highly anisotropic electronic
structure featuring nearly compensated low density electrons
and holes with very small effective mass~\cite{Brandt_1988}. Such an unusual
electronic structure is the basis of unique
electronic properties of other graphitic materials, such as
fullerenes and carbon
nanotubes~\cite{Dresselhaus_Book_1996}, and may lead to
novel manifestations in two-dimensional graphene materials.
For this reason, electron transport in
graphite has recently been the subject of extensive
theoretical~\cite{Khveshchenko_2001, Spataru&Louie_2001}
and experimental~\cite{Kopelevish_2003, Tokumoto&Brooks_2004,
Du&Hebard_2004, Viculis_2003, Lu&Ruoff_1999} investigations.
In particular, interesting size dependent galvanomagnetic effect has been
observed~\cite{Dujardin&Ebbesen_2001, Ohashi_1997+}
in thin layers of graphite, with thickness approaching $\sim$~10~nm.
On this mesoscopic length scale, the electrostatic field-effect (EFE)
modulation of the charge
carrier concentration is expected to be very effective,
owing to the low density of nearly
compensated carriers in graphite. However, the EFE
dependent galvanomagnetic measurements in graphite
have not been carried out in previous studies due
to the difficulty in obtaining adequate sample geometries.

In this letter, we present results from the magnetoresistance (MR)
and Hall resistance measurements in mesoscopic graphite
crystallites consisting of as few as
$\sim$~35~atomic layers. Strong modulation of galvanomagnetic
transport has been observed as the gate electric field changes.
EFE dependent Shubnikov de Hass (SdH)
oscillations, signatures of Landau level formation of electrons and
holes, have been observed at low temperatures. In addition,
the effective mass of electrons
and holes are measured by investigating the temperature damping
of SdH amplitudes for each type of carriers.

\begin{figure}
\includegraphics[width=80mm]{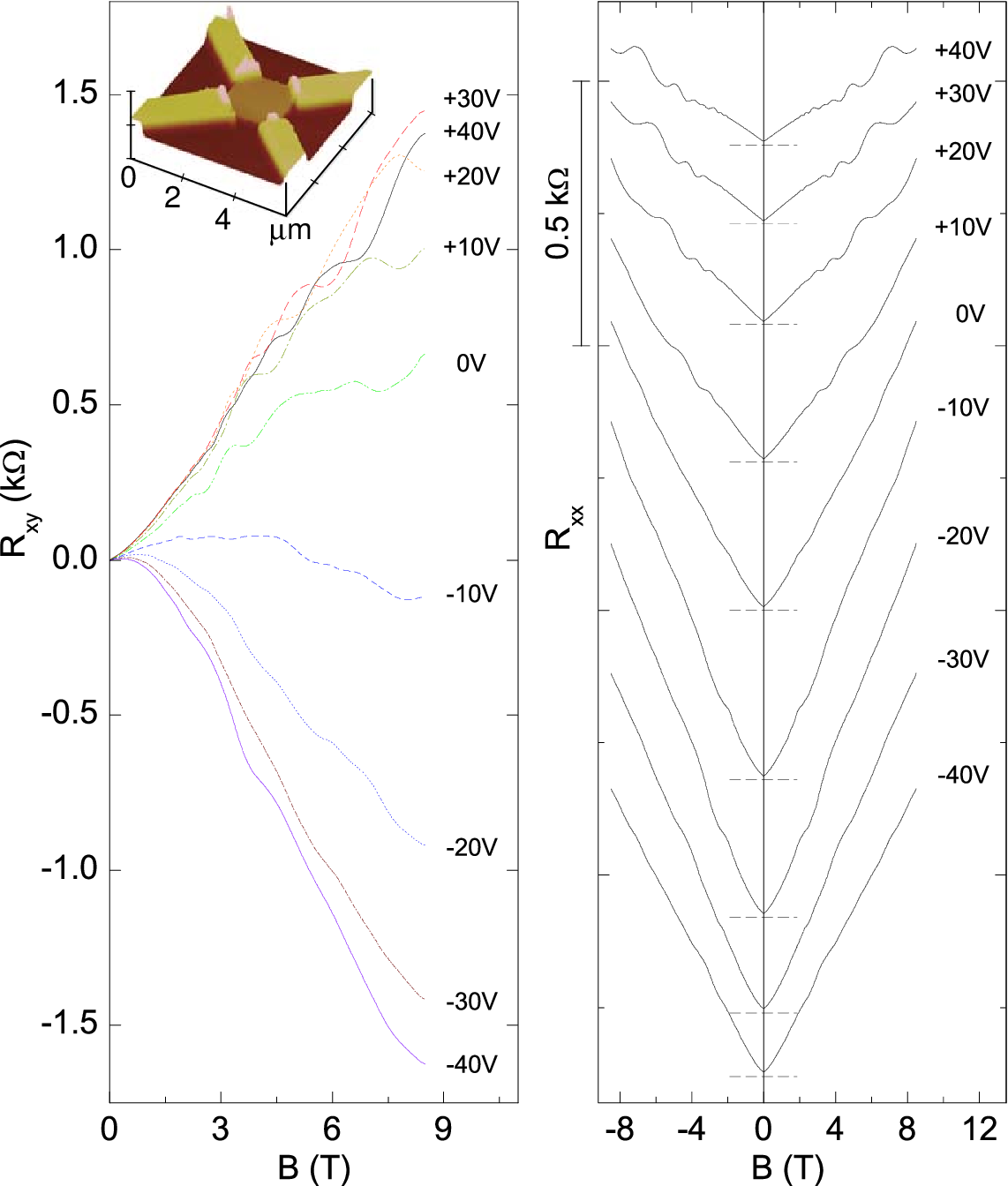}
\caption{The inset shows an AFM image of a 12~nm thick
mesoscopic graphite sample with four electrodes at the corners for
galvanomagnetic measurements. The left and right panel show the
Hall resistance (R$_{xy}$) and magneto-resistance (R$_{xx}$) respectively
as a function
of magnetic field measured at $T=$~1.7~K in this device. Numbers near
each curve indicate the applied gate voltages. In the right panel,
curves are shifted for clarity and the dashed lines correspond to
the zero lines of each curve.} \label{fig1}
\end{figure}

The mesoscopic graphite devices used in this experiment are
fabricated using a unique micro-mechanical method.
The details of the device fabrication are described
elsewhere~\cite{Zhang_APL}. In brief, small graphite crystal
blocks are extracted from bulk highly oriented pyrolytic graphite
(HOPG) using micro-patterning followed by micro-mechanical
manipulation. A detached HOPG block is then transferred
and fixed onto
a micro-machined Si cantilever. By operating an atomic force
microscope (AFM) with load on the graphite mounted cantilever,
very thin layers of graphite crystallites with
lateral size $\sim$~2~$\mu$m and thickness $d $ raging
from 10~-~100~nm are sheared off onto
SiO$_2$/Si substrate. Multiple metal electrodes (Cr/Au) are
then fabricated on the corners, using electron beam lithography
(for AFM image of a typical device, see Fig.~\ref{fig1} inset).
The degenerately doped silicon substrate serves as a
gate electrode with thermally grown
silicon oxide (500~nm) acting as the gate dielectric.

Fig.~\ref{fig1} displays the Hall resistance~($R_{xy}$) and the
MR~($R_{xx}$) as a function of applied magnetic field, $B$,
measured in a 12~nm thick graphite sample at temperature $T=$~1.7~K.
The excitation current is kept at 0.5 $\mu$A for both $R_{xy}$
and $R_{xx}$ measurements.
The magnetic field is applied perpendicularly to the graphite basal plane.
Both quantities exhibit oscillatory features on top of
smooth backgrounds as $B$ varies. Near $V_g\approx$~0~V,
the MR and Hall resistance exhibit similar behaviors to those
observed in high quality bulk graphite~\cite{Soule_1958}.
The ``V" shaped MR background is ascribed to
the general nature of magnetotransport
in materials with coexisting nearly compensated electron and hole
carriers~\cite{Spain_1978}, while the oscillations on top of
the background are related to the SdH
effect, the quantum oscillations due to Landau level formation~\cite{A&M}.
Remarkably, as we vary the gate voltage, $V_g$, the behavior of
$R_{xx}$ and $R_{xy}$ changes dramatically.
The background in the MR is most prominent at
$V_g^{max}\approx$~$-15$~V. As $V_g$ moves away from this value, the
slope of the MR background becomes much smaller. The change of Hall
measurement is even more drastic: $R_{xy}(B)$ changes its sign of
overall slope as $\Delta V_g=V_g-V_g^{max}$ swings from
negative to positive values, indicating that $\Delta V_g$ changes
the dominant majority charge carriers from holes to electrons.
This is a somewhat surprising result at first sight, since
the thickness of the sample (12~nm) is still an order of magnitude
larger than the screening length of graphite
($\lambda_s\approx$~0.4~nm~\cite{Visscher&Falicov}), and thus only
relatively small portion of the sample is affected by the gate electric field. We
will discuss this point quantitatively below.

\begin{figure}
\includegraphics[width=80mm]{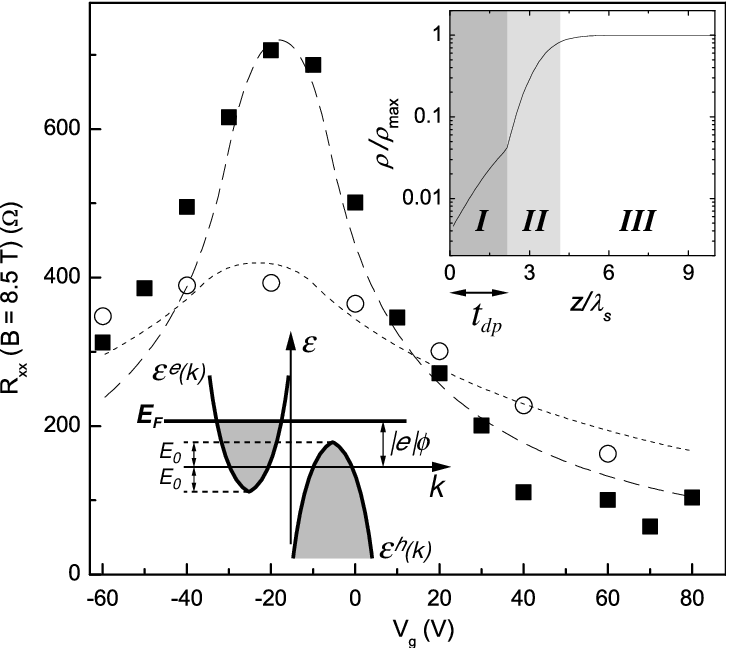}
\caption{Magneto-resistance measured in 12~nm (solid square) and
42~nm (open circle) thick sample at various gate voltages. The
dashed and dotted lines are fits to a model described in text.
The lower inset shows a schematic STB diagram for electron
($\varepsilon^e(k)$) and hole ($\varepsilon^h(k)$) in the
presence of electrostatic potential induced by the gate.
Upper inset represents the local magneto-resistivity across the
sample, assuming $\phi_0=$~8$E_0/|e|$ as an example.
Symbols are defined in text.}
\label{fig2}
\end{figure}

The aforementioned EFE in mesoscopic graphite samples is clearly
presented by observing $R_{xx}$ as a function of gate voltage
at a fixed magnetic field.
Fig.~\ref{fig2} shows $R_{xx}$ as a function of $V_g$ at a large
magnetic field ($B_m=$~8.5~T) for two samples ($d=$~12 and
42~nm)~\cite{footnote1}.
As expected from Fig.~\ref{fig1}, $R_{xx}$ has a peak
near a gate voltage where $\Delta V_g\approx$~0, falling slowly
as $|\Delta V_g|$ becomes large. We found that this gate
dependence strongly depends on $d$. For the 12~nm sample,
$R_{xx}$ is suppressed to $\sim$~10\% of its peak value,
while it is still $\sim$~60\% for the 42~nm sample at $\Delta V_g=$~80~V.
Such a sensitive dependence of $R_{xx}(V_g)|_{B_m}$ on $d$ is
indicative of the reduced EFE by screening of
induced charge near the sample surface.

In order to elucidate the dependence of $R_{xx}$ on $V_g$, we
employ the simple two band (STB) model~\cite{Kelly_1981},
which has been successful
in understanding the MR in graphite~\cite{Tokumoto&Brooks_2004,
Du&Hebard_2004}. The STB model assumes that the bottom of
the electron band and the top
of the hole band overlap with a small band overlap $2E_0$ near
the Fermi energy $E_F$. The resistivity of a sample, $\rho$,
in the presence of a magnetic field can be
expressed by~\cite{Spain_1978}:
\begin{equation}
\label{eq_1}
\frac{\Delta\rho}{\rho_0}=\frac{4\mu^2B^2n_en_h/(n_e+n_h)^2}{1+[\mu
B(n_e-n_h)/(n_e+n_h)]^2}
\end{equation}
where $\rho_0=\rho(B=0)$, $\Delta\rho=\rho(B)-\rho_0$, $\mu$ is
the average carriers mobility, and $n_e$ and $n_h$ are the carrier
concentrations of electrons and holes, respectively. Generally,
$\Delta\rho$ varies the most as a function of $B$ when electrons
and holes are nearly compensated~(i.e., $n_e\approx n_h$).
From Fig.~\ref{fig2}, we infer that this condition is met
at $V_g\approx V_g^{max}$ where the
growth of the MR background as a function of $B$ is strongest in our samples
(see the curves for $V_g$~=~$-10$~V and $V_g$~=~$-20$~V
in Fig.~\ref{fig1})~\cite{footnote2}.
As $\Delta V_g$ increases from zero, the induced charge in the sample screens
the gate electric field and the electrostatic potential in the sample is given
by $\phi(z)=\phi_0 e^{-z/\lambda_s}$, where $z$ is measured from
the interface between the
sample and the substrate. The constant $\phi_0$ can be determined
from the electrostatic gate coupling to the sample.
By integrating over the induced charge in the
sample, we obtain $\phi_0=\alpha\Delta V_g$ with
the constant
$\alpha^{-1}=1+\varepsilon_0(1-e^{-d/\lambda_s})/\lambda_sC_g$,
where $C_g$ is the gate capacitance per unit area of the sample
and $\varepsilon_0$ is the vacuum permittivity~\cite{Zhang_APL}.

We incorporate this local electrostatic potential to the STB model
by considering a gradient in $n_e$ and $n_h$.
Suppose $\Delta V_g>$~0, the local electrostatic
potential will pull down the electron and hole bands by
$|e|\phi(z)$ (Fig.~\ref{fig2} lower inset).
For a sufficiently
large gate voltage, such that $|e|\phi_0>E_0$, the sample can be devided
into three regions by introducing a hole depletion depth,
$t_{dp}=\lambda_s\log(|e|\phi_0/E_0)$: (I)~$0<z<t_{dp}$, where
$n_h\simeq$~0; (II)~$t_{dp}<z\lesssim t_{dp}+\lambda_s$, where
$0<n_h<n_e$; and (III)~$z>t_{dp}+\lambda_s$, where $n_h\approx
n_e$. In region (I), only electrons participate in the
transport, and $\rho(z)$ increases as $z$
approaches zero, owing to the electric field induced accumulation of $n_e$ near
the surface. In region (II), the MR is described by Eq.~\ref{eq_1}, so
a steep increase of $\rho(z)$ is expected as
$n_e-n_h$ becomes small. In region (III), the gate electric field
is completely screened, so $n_e\approx n_h$ and
$\rho(z)\approx\rho_{max}=R_{xx}(\Delta V_g=0)d$.
The exact opposite
argument works for a sufficiently large negative gate voltage,
where electrons are depleted.
Note that for small $|\Delta V_g|$, where $|e\phi_0|<E_0$, region (I)
disappears (i.e., $t_{dp}=0$). From above discussions, we now build a
quantitative model to describe $\rho(z)$. According
to the STB model
$n_e(\epsilon)$, $n_h(\epsilon) \propto \epsilon^{3/2}$, where
$\epsilon$ is measured from the bottom of the respective band
edge, $\rho(z)/\rho_{max}$ is obtained from Eq.~\ref{eq_1}
(Fig.~\ref{fig2} upper inset). Then the
resistance of the sample can be evaluated from
$R^{-1}=\int_0^d\rho^{-1}(z)dz$ for fixed $B$ and $V_g$.
Following in this way, a reasonable
fit is obtained for both 12~nm (dashed line) and 42~nm (dotted
line) samples as shown in Fig.~\ref{fig2}. In this fit, we use
$E_0=$~15~meV, a value quoted in~\cite{Brandt_1988}, and obtain
$C_g=$~26~aF/$\mu$m$^2$~(12~nm sample) and
24~aF/$\mu$m$^2$~(42~nm sample) as a result of the fit.
These capacitance values are in reasonable agreement with
our previous estimations in a different
analysis on the same samples~\cite{Zhang_APL}. It is noteworthy that
for a large $V_g$ such that $|e|\phi_0\gg E_0$, $\rho(z)\ll \rho_{max}$
in region (I), and thus a significant portion of the total current flows
in this region. Furthermore, as $\Delta V_g$ increases, $t_{dp}$
grows only logarithmically. Even at $\Delta V_g\approx$~100~V, the
largest gate voltage applied, $t_{dp}\approx$~1~nm, which
corresponds to only $\sim$~3 bottom layers.  Therefore, only a few of the
bottom layers of the sample are responsible for the
observed EFE modulation
of the galvanomagnetic transport quantities.

\begin{figure}
\includegraphics[width=80mm]{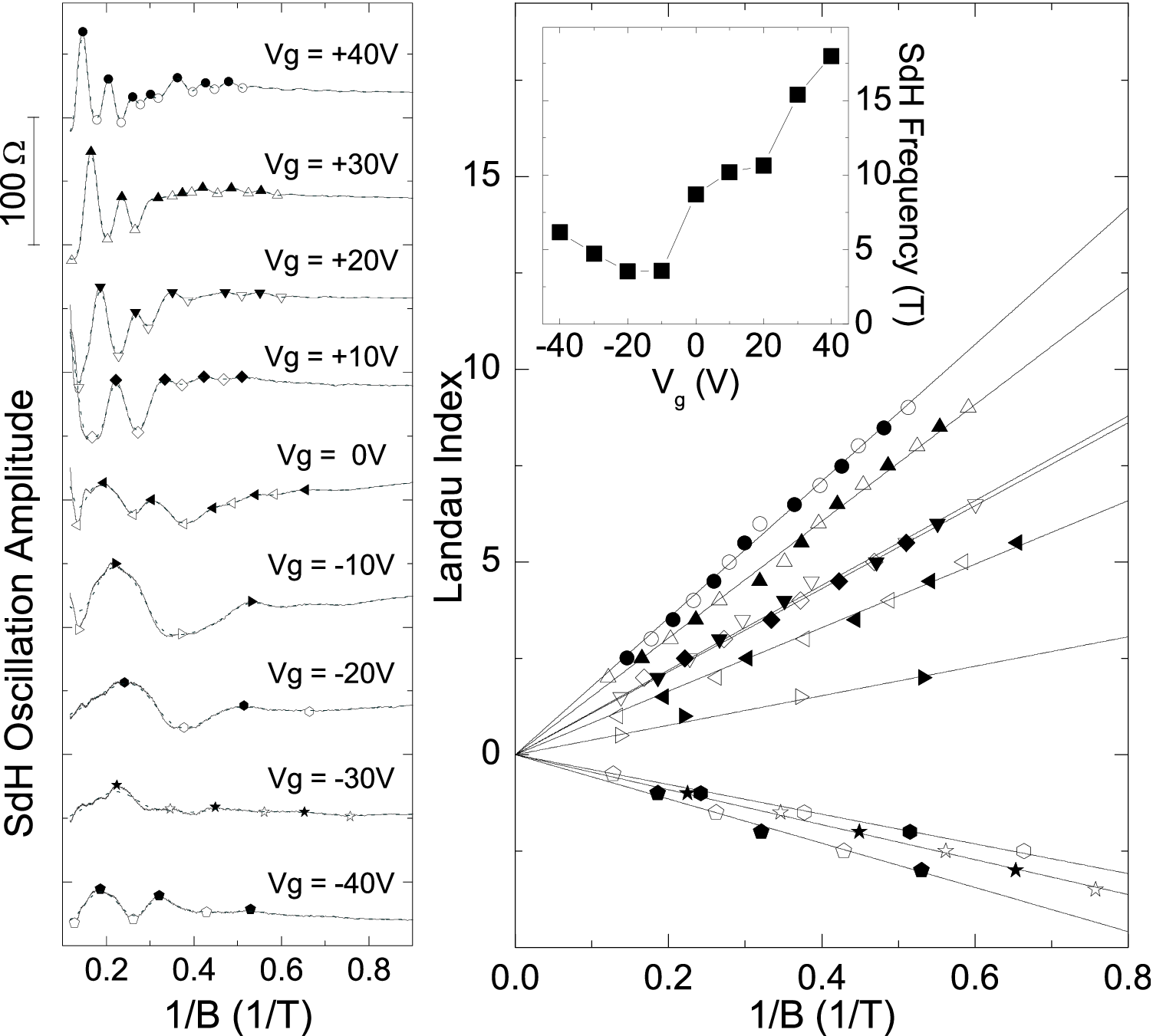}
\caption{(a) The SdH Oscillations observed in
Fig.~\ref{fig1}, after subtraction of smooth backgrounds.
Solid (open) symbols correspond to peak (valley) of
the oscillations found after passing the curve through a low pass
filter (dotted line). Curves are displaced for clarity.
(b) Landau plots (see text) obtained from (a). Negative
indices are assigned to the hole branch for clarity. Lines are linear
fits to each set of points at different $V_g$. Inset: the frequency
of the SdH oscillations obtained from the slopes of the line fits in
(b) as a function of gate voltage.} \label{fig3}
\end{figure}

We now turn our attention to the quantum oscillations observed in
our mesoscopic graphite samples.
The strong EFE modulation of the carrier density in the bottom layers
allows us to probe the quantum oscillations in these layers
with a continuously tunable carrier concentration.
Fig.~\ref{fig3}(a) redisplays the separated SdH oscillations as a function of
$B^{-1}$, obtained from the MR data shown in Fig.~\ref{fig1} after
subtracting out the smooth background.
The SdH oscillations indicate the oscillatory
density of states at $E_F$ as a quantized Landau level passes
through $E_F$. The frequency of SdH oscillations, $f_s$, is related
to the extremal area of the electron and hole pockets of the Fermi
surface by $f_s=\hbar cA^{e, h}_k/2\pi |e|$, where $A^e_k$ and $A^h_k$ are the
areas of extremal electron and hole pockets, and $\hbar$ and $c$ are
Plank constant and speed of light respectively~\cite{A&M}. Since
$n_e$ and $n_h$ are modulated by $V_g$, the observed variation of
$f_s$ can be explained by the change of $A^{e, h}_k$.

In order to demonstrate the change of $f_s$ quantitatively, we
first locate the major peaks (solid symbols) and valleys (open
symbols) in the SdH oscillations after low pass filtering of the
data~\cite{footnote3}. The value of $B^{-1}$ for a peak (valley), $B_m^{-1}$,
is indexed by $\nu$, an integer (a half integer) number that corresponds to
the Landau level responsible for the particular oscillation.
Fig.~\ref{fig3}(b) shows that each set of points ($B_m^{-1}, \nu$)
at a given $V_g$ are
on a straight line that intercepts the origin, implying that the period of SdH
oscillations is regular. From the slope of these lines we obtain
$f_s$ at different $V_g$ (Fig.~\ref{fig3}(b) inset). The obtained
$f_s$'s are increasing with $|\Delta V_g|$. Therefore, we
believe that the obtained $f_s$ corresponds to $A^e_k$
for $\Delta V_g>$~0 and to $A^h_k$ for $\Delta V_g<$~0.
This conclusion allows us to compare the experimentally observed $f_s$
with the expected value from the STB model.
Assuming the Fermi surface of graphite is described by the overlap
of electron and hole bands in STB model,
$A^e_k\propto (\alpha\Delta V_g +E_0)$ and $A^h_k\propto
(-\alpha\Delta V_g +E_0)$. This relationship leads to
$f_s(|\Delta V_g|)/f^0_s=1+\alpha|\Delta V_g|/E_0$, where
$f^0_s=f_s(\Delta V_g=0)$. From the values of $\alpha$ and $E_0$,
determined separately above, we
estimate $f(\Delta V_g=50 V)/f_0 \approx$~4.8, which is in
reasonable agreement with the experimentally observed ratio~4.3.

\begin{figure}
\includegraphics[width=80mm]{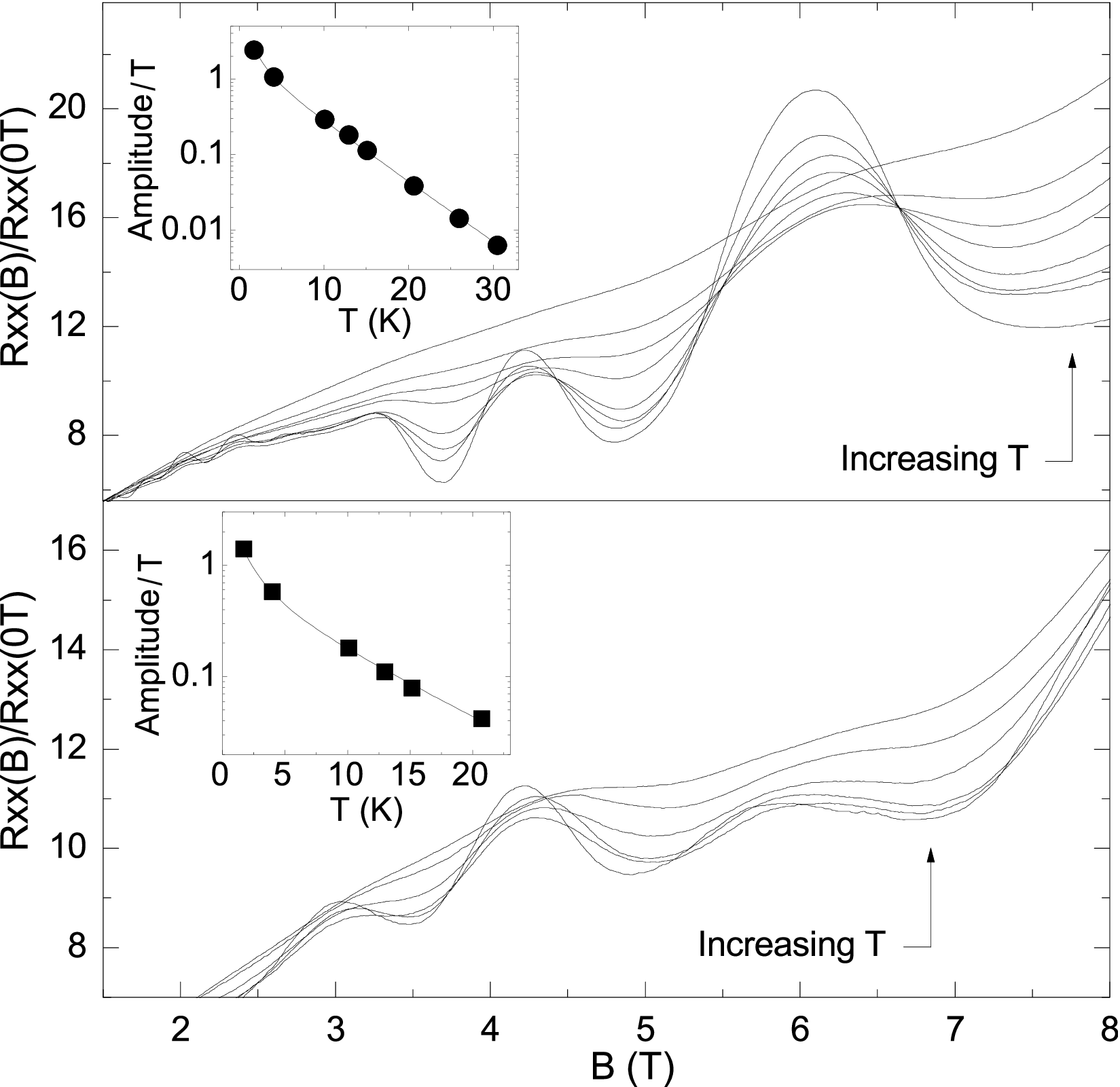}
\caption{Normalized magneto-resistance of the sample in Fig~\ref{fig1}
at $V_g=$~40~V (upper panel) and $V_g=-$~60~V (lower panel). For
the upper panel data are taken at temperatures 1.7, 4, 10, 13, 15, 20,
25, and 30~K. For the lower panel, data are taken at temperatures
1.7, 4, 10, 13, 15, and 20~K.
Insets: SdH oscillation amplitudes divided by temperature, $T$,
at a fixed magnetic field
at above temperatures. The solid lines are fits to a model
(see text).} \label{fig4}
\end{figure}

Finally, we discuss the temperature dependence of the SdH
oscillations. Fig.~\ref{fig4} shows the oscillatory MR at two
extreme gate voltages, $V_g=+$40~V and $V_g=-$60~V, at various
temperatures. At these extreme gate voltages, the transport in the sample
is dominated by only one type of carriers in a few
bottom layers. Thus, the SdH oscillations in the upper
(lower) panel of the figure correspond to electron
(hole) Landau levels in the sample. In both cases,
the observed SdH oscillation amplitude is gradually damped away as
the temperature increases.
The temperature dependent
SdH oscillation amplitude has been used to extract the effective mass of
charge carriers~\cite{SdH&MASS}.
At a fixed magnetic field, the temperature damping factor of the SdH
oscillation amplitude is given by:
\begin{equation}
\label{eq_2}
R_{T}=\frac{2\pi^2k_BTm^*/e\hbar B}{\sinh(2\pi^2k_BTm^*/e\hbar B)}
\end{equation} where $m^*$ is the effective mass
of the carriers. We find that Eq.~\ref{eq_2}
fits the observed amplitude damping very well (Fig.~\ref{fig4} insets).
As a result from the fittings,
the effective electron mass $m^*_e=$~(0.052$~\pm$~0.002)$m_e$
and hole mass $m^*_h=$~(0.038~$\pm$~0.002)$m_e$ are obtained, where
$m_e$ is the bare electron mass. These values agree well with
0.057$m_e$ and 0.039$m_e$, reported effective mass of electrons and holes
in high quality bulk graphite crystal~\cite{Brandt_1988}.

In summary, we report galvanomagnetic transport in mesoscopic
graphite samples consisting of tens of graphene layers. Strong
modulation of the Hall resistance as well as the magneto-resistance
has been observed as the applied gate voltage changes. The Landau
level formation of electron and hole carriers is also tuned by
the gate. The unique experimental method discussed here can be
applied to other layered materials to investigate novel transport
phenomena in unconventional two dimensional systems.

We thank H. L. Stormer, A. Millis and I. Aleiner for helpful
discussions. This work is supported primarily by the Nanoscale
Science and Engineering Initiative of the National Science
Foundation under NSF Award Number \mbox{CHE-0117752} and by the
New York State Office of Science, Technology, and Academic
Research \mbox{(NYSTAR)}.


\begin{references}

\bibitem{Brandt_1988}N. B. Brandt, S. M. Chudinov, and Y. G.
Ponomarev, Semimetals 1: Graphite and its compunds (North-Holland,
1988).

\bibitem{Dresselhaus_Book_1996}M. S. Dresselhaus, G. Dresselhaus,
and P. C. Eklund, {\it Science of Fullerenes and Carbon Nanotubes}
(Academic, 1996).

\bibitem{Khveshchenko_2001} D. V. Khveshchenko, Phys.
Rev.Lett., {\bf 87}, 246401 (2001); D. V. Khveshchenko, {\it
ibid}, {\bf 87}, 246802 (2001).

\bibitem{Spataru&Louie_2001} C. D. Spataru, M. A. Cazalilla, A. Rubio,
L. X. Benedict, P. M. Echenique, and S. G. Louie, Phys. Rev. Lett.,
{\bf 87}, 246405 (2001).

\bibitem{Kopelevish_2003} Y, Kopelevish, J. H. S. Torres,
R. R. da Silva, F. Mrowka, H. Kempa, and P. Esquinazi, Phys.
Rev.Lett., {\bf 90}, 156402 (2003).

\bibitem{Tokumoto&Brooks_2004} T. Tokumoto, E. Jobiliong, E. S. Choi, Y. Oshima, and J. S. Brooks,
Solid. State. Commun., {\bf 129}, 599 (2004).

\bibitem{Du&Hebard_2004} X. Du, S. Tsai, D. L. Maslov, and A. F. Hebard, cond-mat/0404725 (2004).

\bibitem{Viculis_2003} L. M. Viculis, J. J. Jack, and R. B. Kaner,
Science {\bf 299}, 1361 (2003).

\bibitem{Lu&Ruoff_1999} X. Lu, H. Huang, N. Nemchuk, and R. Ruoff,
Appl. Phys. Lett. {\bf 75}, 193 (1999); X. Lu, M. Yu, H. Huang,
and R. Ruoff, Nanotechnology {\bf 10}, 269 (1999).

\bibitem{Dujardin&Ebbesen_2001}  E. Dujardin, T. Thio, H. Lezec, and T. W. Ebbesen,
Appl. Phys. Lett. {\bf 79}, 2474 (2001).

\bibitem{Ohashi_1997+}Y. Ohashi, T. Hironaka, T. Kubo, and K. Shiiki,
Tanso {\bf 1997}, 235 (1997); Y. Ohashi, T. Hironaka, T. Kubo, and
K. Shiiki, Tanso {\bf 2000}, 410 (2000).

\bibitem{Zhang_APL} Y. Zhang, J. P. Small, W. Pontius, and P. Kim,
submitted for publication.

\bibitem{Soule_1958} D. E. Soule,
Phys. Rev. {\bf 11}, 698 (1958).

\bibitem{Spain_1978} I. L. Spain, Carbon {\bf 17}, 209 (1978).

\bibitem{A&M} N. W. Ashcroft and N. D. Mermin,
{\it Solid State Physics} (Harcourt College,1976).

\bibitem{Visscher&Falicov}P. R. Visscher and L. M. Falicov,
Phys. Rev. B {\bf 3}, 2541 (1971).

\bibitem{footnote1} We measured 6 mesoscopic graphite samples
with the thickness ranging 12~-~95~nm. Similar result was found
the sample with the similar geometry.

\bibitem{Kelly_1981} For an extensive summary see, for example, B. T. Kelly
{\it Physics of Graphite} (Applied Science,1981), pp285.

\bibitem{footnote2} $\Delta R_{xx}(H)/R_{xx}$ increases rather linearly than
quadratic increase as predicted from the STB model. Similar behavior has
been observed other bulk graphite samples (see, for e.g., Y.
Kaburagi and Y. Hishiyama, Carbon {\bf 33}, 1505 (1995) and
reference therein.). This deviation is, however, not essential in
our argument in this paper.

\bibitem{footnote3} In principle, we expect to see two distinct
oscillation periods caused by electrons and holes in
graphite. However, due to the finite lateral size of sample
($L=$2~$\mu$m), only the cyclotron orbits whose diameters are much
smaller than $L$ contribute most in the observed SdH oscillations.

\bibitem{SdH&MASS} D. Shoenberg, {\it Magnetic oscillations in metals},
(Cambridge University Press, 1984).


\end{references}
\end{document}